\documentclass[twocolumn]{article}
\usepackage{times}
\usepackage{mathptmx}
\usepackage{multirow}
\usepackage{hyperref}
\usepackage[utf8]{inputenc}
\usepackage[T1]{fontenc}
\usepackage{amsmath}
\usepackage{textcomp}
\usepackage[top=2.5cm,bottom=2.5cm,left=2.5cm,right=2.5cm]{geometry}
\usepackage[sort&compress]{natbib}
\usepackage{multicol}
\usepackage{subcaption}
\usepackage{float}
\usepackage[affil-it]{authblk}
\setcitestyle{numbers,square,comma}

\usepackage{stfloats}

\usepackage[color=blue!10,textsize=footnotesize,textwidth=25mm]{todonotes}

\providecommand{\keywords}[1]{\textbf{Keywords:} #1}
\usepackage{enumitem}
\setlist{topsep=1ex,itemsep=0.5ex,parsep=0pt,partopsep=0pt}  

\newcommand{\thePurpose}{Machine Learning-Based Detection of MCP Attacks} 

\title{\thePurpose}

\author[1]{Tobias Mattsson}
\author[1]{Samuel Nyberg}

\author[2]{Anton Borg\footnote{Corresponding author: anton.borg@bth.se}}

\author[3,4]{Ricardo Britto}
\affil[1]{Blekinge Institute of Technnology, Sweden}
\affil[2]{Department of Computer Science, Blekinge Institute of Technology, Sweden}
\affil[3]{Ericsson, Sweden}
\affil[4]{Department of Software Engineering, Blekinge Institute of Technology, Sweden}

\begin{document}

\maketitle
\begin{abstract}

The Model Context Protocol (MCP) is a new and emerging technology that extends the functionality of large language models, improving workflows but also exposing users to a new attack surface. Several studies have highlighted related security flaws, but MCP attack detection remains underexplored. To address this research gap, this study develops and evaluates a range of supervised machine learning approaches, including both traditional and deep-learning models. We evaluated the systems on the detection of malicious MCP tool descriptions in two scenarios: (1) a binary classification task distinguishing malicious from benign tools, and (2) a multiclass classification task identifying the attack type while separating benign from malicious tools. In addition to the machine learning models, we compared a rule-based approach that serves as a baseline. The results indicate that several of the developed models achieved 100\% F1-score on the binary classification task. In the multiclass scenario, the SVC and BERT models performed best, achieving F1 scores of 90.56\% and 88.33\%, respectively. Confusion matrices were also used to visualize the full distribution of predictions often missed by traditional metrics, providing additional insight for selecting the best-fitting solution in real-world scenarios. This study presents an addition to the MCP defence area, showing that machine learning models can perform exceptionally well in separating malicious and benign data points. To apply the solution in a live environment, a middleware was developed to classify which MCP tools are safe to use before execution, and block the ones that are not safe. Furthermore, the study shows that these models can outperform traditional rule-based solutions currently in use in the field.

\end{abstract}


\section{Introduction}
\label{sec:intro}
In today's era of AI and automation, a new technology, Model Context Protocol (MCP) \cite{MCP} has emerged. With MCP servers and tools, the possibility of adding additional context to pretrained Large Language Models (LLM) can extend the functionality of chatbot agents like ChatGPT and Claude. Since MCP servers and tools improve the already impressive capabilities of AI agents, the field has been evolving rapidly during the last year. Companies are increasingly integrating MCP tools into their workflows to enhance employee efficiency. Many invest substantial resources into developing tailored tools for company-specific needs or adopting existing ones.

With the rapid adoption of MCP tools and their widespread use by organizations, it is necessary to raise questions about their security. Since the technology is still relatively new, published in late 2024 by Anthropic \cite{MCP}, most users are unaware of the potential security vulnerabilities that can lurk within, such as Tool Poisoning Attacks (TPA), rug pulls, and file manipulation, among others. All of which can lead to severe leakage of a company's or an individual's confidential data, harm caused to systems, and breach of trust towards benign tools. The problem lies in the difficulty of validating tools that follow the MCP protocol, and in the ease of interchanging tools, which can lead to accidental installations with the intent to harm. Together with a rapid increase in documented attack variants over the past year \cite{multipleAttacks}. Although vulnerabilities have been researched and categorized, defensive countermeasures remain understudied.

Even though this domain is new and minimal research has been conducted, it shares many similarities with email phishing detection \cite{MachineLearningAlgorithmForDetectingSuspiciousEmail}. At an abstract level, both domains operate on natural language data and identify if a sequence of text matches any flagged patterns. Given the similarities between the fields, much knowledge can be drawn from research on email phishing detection and applied to the MCP attack detection domain.

Despite these promising results in related domains, defensive techniques for MCP-based attacks remain largely underexplored. Thus, revealing a significant gap in the research on defensive methods against these vulnerabilities, this paper aims to develop a deep learning-based detection tool that analyzes the MCP connection by combining established industry detection standards with an innovative approach, previously limited to theoretical research or to other domains.

The main objective of this research is to develop a validation method for MCP tools. The original design envisioned a multi-layered defense that combined complementary detection techniques: publicly available rule-based systems as first-line filters and a novel supervised machine-learning layer as the core defensive component. In the study, state-of-the-art rule-based detectors served as baseline comparators, while the main contribution focused on designing and evaluating a new ML-based approach not previously explored in this domain. Empirical results, however, demonstrated that the machine-learning models substantially outperformed the rule-based baselines, motivating a standalone ML-centric validation method.





\section{Related Work}
The field of agentic AI and MCP is expanding rapidly. This growth increases the attack surface and exposes more actors to vulnerabilities. Therefore, understanding the current state of research in this area is essential to ensure this work incorporates the latest findings. This section provides an overview of articles and resources that have been important and relevant to this work.

Rule-based methods have been widely studied in early email spam and phishing detection systems \cite{NaiveBayesianAndKeyword-BasedAnti-Spam}. These approaches rely on manually defined heuristics such as keyword lists, header features, and structural patterns, and have been shown to be effective for identifying well-known and relatively static spam types. Later work introduced machine learning (ML) approaches, which learn statistical and semantic patterns rather than relying on fixed rules. Prior work shows that ML methods improved detection performance on unseen phishing attacks \cite{MachineLearningAlgorithmForDetectingSuspiciousEmail}. Despite these advantages, several works note limitations of purely ML-based systems, including sensitivity to the quality of the training data, sampling bias, and increased false-positive rates. To address these issues, hybrid detection frameworks, often referred to as neuro-symbolic approaches, have been explored, integrating rule-based reasoning with ML models. Results from the spam and phishing literature indicate that such hybrid systems can improve robustness and precision compared to using either rule-based or ML-based methods alone \cite{aHybridMachineLearningApproachforSecuringEmails}.

Yongjian Gou \textit{et al.} (2025) \cite{multipleAttacks} acknowledge the lack of research done on the different types of attacks on MCP. To address this gap, they present a comprehensive analysis of 31 known attack types. The result is shown in a table, where the attacks are categorized and assigned an individual attack efficiency score based on how frequently they succeeded. Furthermore, researchers provide a collection of the attacks in what they call the MCP Attack Library (MCPLib). The contribution of MCPLib is highly relevant to this study, as it provides a solid foundation for identifying and constructing effective training data.

Zhiqiang Wang \textit{et al.} (2025) \cite{MCPTox} presents a study on a benchmark called MCPTox. The benchmark is designed to systematically evaluate agent robustness against Tool Poisoning attacks. It is constructed from 45 authentic MCP servers and 353 tools. From these authentic MCP tools, the researchers crafted effective poisoned tool descriptions to give each MCP server multiple malicious test cases. To demonstrate TPAs' stealth capabilities, the malicious actions were never performed by the poisoned tools; instead, they indirectly influenced benign tools to perform them. The benchmark was evaluated on 20 of the most widely adopted LLMs. The results indicated an average Attack Success Rate (ASR) of 36.5\%, while some of the more powerful models achieved an ASR of 72.8\%. Based on the benchmark, the researchers concluded that widespread vulnerabilities exist in popular LLMs and that a pre-execution security mechanism is needed to safeguard against these attacks.

The study by Kingshuk Debnath \textit{et al.} (2022) \cite{emailSpamDetection} focused on email spam detection using both traditional machine learning classifiers and deep learning models, including LSTM, BiLSTM, and BERT. The experiments were conducted using the Enron email dataset, which consists of 5,171 emails categorized into two classes: ham and spam. The study demonstrated that transformer models consistently outperformed classical machine learning methods, with BERT obtaining the highest accuracy of 99.14\%. These results highlight the effectiveness of deep learning models in a similar NLP domain and provide an indication of their potential performance in this study.

 From the limited research that has previously been done, they have characterized MCP attacks, produced some datasets of known incidents, and identified the need for pre-execution detection. In addition, research on email spam detection indicates that deep learning models show great effectiveness in a field similar to that of MCP attack detection. These factors together indicate a meaningful research gap that has not yet been fully explored, highlighting the relevance of this work.

\section{Research Design}

This section describes the methodology used to design and conduct the experiments presented in this paper, including the chosen research design, dataset preparation and exploration, and the supervised models employed along with their configurations.

The methodology chosen for this research paper is DSM, as it aligns well with combining industry and academia. Specifically, this research will follow the process proposed by Offerman et al. \cite{inproceedings}, and applied as described by C.Wohlin and P.Runeson \cite{WOHLIN2021106678}. DSM follows 11 steps, split into three distinct phases. It utilizes a sequential mixed-methods approach \cite{creswell2009research}, combining qualitative and quantitative approaches. Qualitative research is used in the early stages to understand and shape the problem in its real-world context, while quantitative methods are applied in later stages for development, evaluation, and iterative improvement.

First, the process undergoes a problem identification phase, during which a literature review is conducted to assess topic relevance across academia and industry, identifying the problem and its underlying cause.

Phase two consists of solution design, often seen as the development phase. It usually begins with a second, more in-depth literature review of previously existing solutions and state-of-the-art research, aiming to identify gaps and possible solutions to the proposed problem. During the second stage, more knowledge about the problem is acquired, and the problem may need revision, iterating back to phase one. The primary goal of phase two is to define the \textit{artifact}, which, according to C. Wohlin and P. Runeson \cite{WOHLIN2021106678}, is a tool, model, or technique created to answer the research questions or hypotheses.

The third phase consists of evaluating the \textit{artifact}, assessing the problem, and determining whether the research question has been properly addressed or whether the process needs to return to phase two for further refinement and analysis.

In choosing a sequential mix of research methodologies, this study adopts DSM, which has several advantages for this type of research over other well-studied methods \cite{thesisproijectguide}, such as case study \cite{casestudy}, implementation \cite{implementationscience}, or experimental methods\cite{experimentalmethods}, focusing on either qualitative or quantitative research. The case study approach focuses on the collection and analysis of data and processes. Which would be a viable method if the aim were to examine different existing attacks. In contrast, the objective of this study is to develop a new defense method to counter existing attacks. Therefore, a purely qualitative method is unsuitable for this research, as it does not facilitate the development of new solutions or an iterative process.

Implementation, a method commonly used in computer science research when developing new software architectures, algorithms, or programs, also known as \textit{artifacts} \cite{WOHLIN2021106678}, is created to improve existing approaches. Such \textit{artifact} typically requires comparison with existing techniques, which will be difficult in the current study because the technique is state-of-the-art and has no direct predecessors.

Experimentation, one of the fundamental scientific research methods, aims to verify or falsify a hypothesis by examining variables and how they are affected by different experimental conditions. Experimentation methods, which align to some degree with the experimental nature of this project, are especially useful when developing artificial intelligence models and verifying results. However, this study focuses on creating a malicious Trojan detection tool, which does not fully conform to the practice of hypothesis testing due to the iterative process required for development and the creation of the \textit{artifact}.

DSM enables us to combine the best elements of several well‑established methods, allowing components of previously described research methods to be reflected in the DSM. It improves applicability when developing an \textit{artifact} intended for both academia and industry. Further, it supports an iterative process that allows refinement of the \textit{artifact} between evaluation stages, yielding a well‑tested product that, when applied in industry and research, is reproducible, validated, and enables further academic improvements.

\subsection{Research flow}
The chosen research flow for this project is inspired by the DSM methodology and follows 11 stages, divided into three phases. The graph below illustrates the steps of the research flow and logic behind them.
\begin{figure}[H]
    \centering
    \includegraphics[width=1\linewidth]{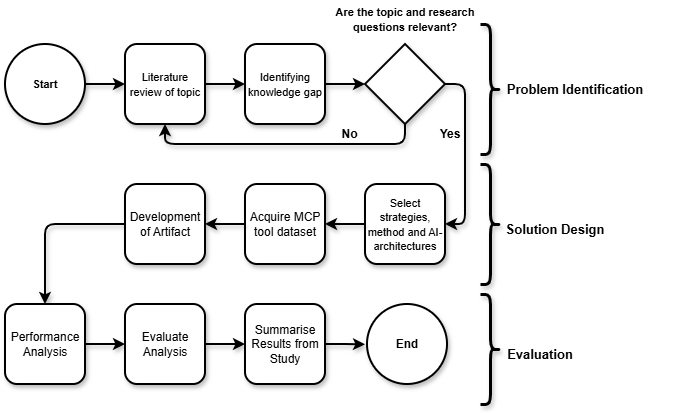}
    \caption{Design flow of project structure}
    \label{fig:DSMFlow}
\end{figure}

\subsection{Problem Identification}
The initial phase of a research project involves acquiring a comprehensive understanding of the relevant field, grasping existing knowledge, and identifying knowledge gaps within. To achieve this, a literature review will be performed focusing on the most relevant and recently published articles on the subject. This step is crucial to getting a clear and accurate understanding of the research problem. Once sufficient knowledge of the field has been acquired, a gap in academia has been identified, the project scope defined, and research questions formulated.

\subsection{Solution Design}
\label{solution_design}
The second phase of the design solution methodology consisted of constructing and developing the \textit{artifact}. To achieve the goal of this research and reduce the identified knowledge gap, an MCP-tool connection analytics tool was created, featuring machine-learning models to analyze MCP-tool descriptions. This phase required multiple steps, from collecting and preparing data to developing a rule-based analysis pipeline as a baseline for benchmarking performance to deciding which machine-learning models to adapt.

During this study, we framed the same data under two classification settings: (1) binary classification of MCP tools into malicious or benign, and (2) multi-class classification with eleven classes—ten distinct tool-injection attacks and one benign class. The binary (1) case was created by treating all malicious classes as a single label; the binary classification task is the fundamental concept of any malware detection, distinguishing between benign and malicious tools. Setting (2) served as an extension of the detection tool, allowing users to identify the type of attack they were encountering.

\subsubsection{The Artifact}
\begin{figure*}[b]
    \centering
    \includegraphics[width=1\linewidth]{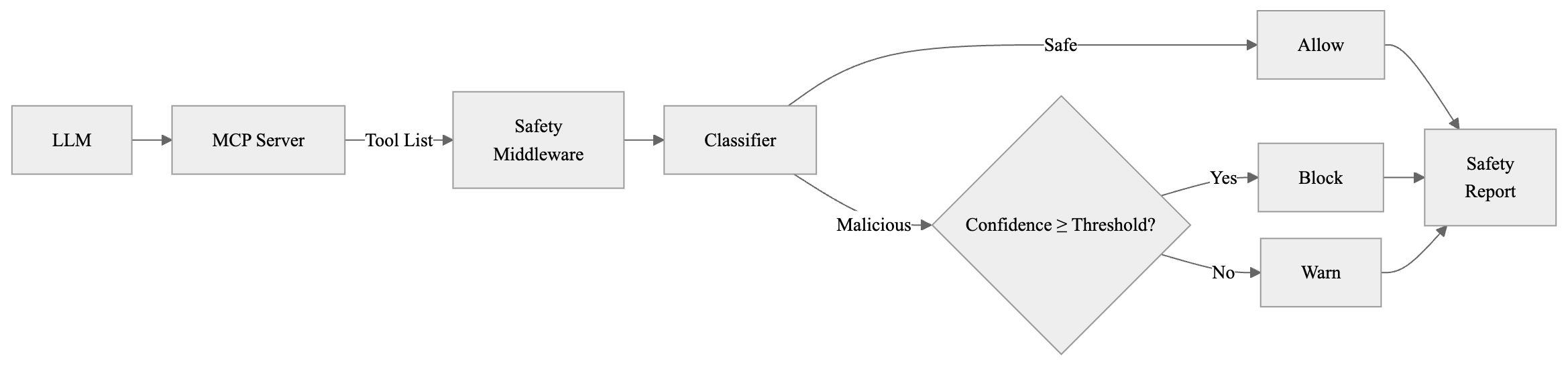}
    \caption{Flowchart for middleware when intercepting MCP-tool call}
    \label{fig:middleware}
\end{figure*}

After testing and evaluating the models presented in this study, the findings will be deployed in a live environment by developing the \textit{artifact}. The \textit{artifact} consists of one of the top-performing machine learning models encapsulated in a script. It accepts MCP tool descriptions and classifies their content to determine whether the intent is benign or malicious. The script is invoked by LLMs and runs as a pre-MCP-tool execution hook before all MCP-tool calls. Enabling the \textit{artifact} to classify the MCP tools that have been selected by the LLM, generate a report for the user, and then intercept the call, allowing or blocking the tool before execution. The \textit{artifact} follows the idea of a middleware, which can intercept LLM calls, but unlike capability-oriented \cite{mcp_middleware}, which usually serves as a translation between the LLM-agent and the environment, the \textit{artifact} developed for this study acts as a firewall between the LLM-agent and the tool. The flow for the \textit{artifact} is shown in Figure \ref{fig:middleware}.


\subsubsection{Data Collection}
\label{sec:aquiring_data}
To apply supervised machine learning models~\cite{SupervisedLearning}, all data points were required to be annotated with appropriate labels. In this case, the task involved binary classification, distinguishing between malicious data points (MCP tool instructions containing embedded tool-poisoning attacks) and benign data points (regular tool descriptions without attacks). Data for this study were collected through two distinct approaches. 

For malicious data points, a dataset composed solely of malicious tool descriptions from the study \cite{MCPTox} was used, which provides an analytical review of various MCP tool-poisoning attacks. The dataset contains 485 malicious tool descriptions across 11 classes.

\begin{figure}[H]
    \centering
    \includegraphics[width=0.75\linewidth]{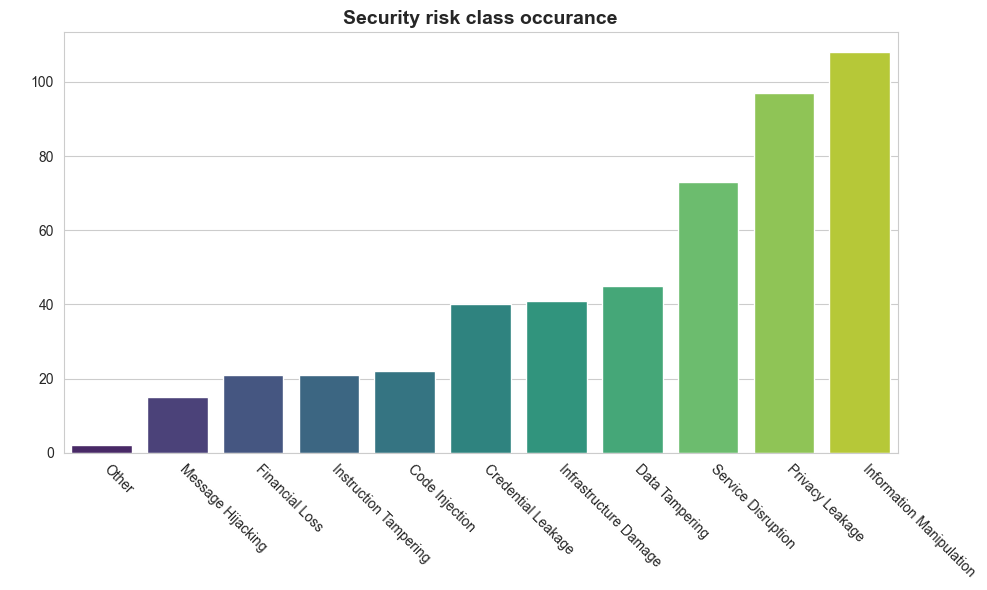}
    \caption{Class distribution of malicious MCP-tox dataset}
    \label{fig:placeholder}
\end{figure}

Benign data were gathered using a custom MCP tool repository crawler and scraper, built specifically for this project, which sources tool descriptions from various GitHub registries, such as \cite{MCPRegistry}. Tool descriptions were obtained by scanning MCP tool repositories for Python, TypeScript, and JavaScript files, as these were identified as the most common programming languages used for MCP tool development. Once these file types were detected, a custom parser searched for 12 predefined tool-definition patterns to locate the sections where MCP tool descriptions were written. These patterns reflected different architectures and tool description protocols.

For example, common patterns and decorators such as \textit{@server.tool()} and \textit{server.tool()} are used in tools built on the FastMCP framework \cite{FastMCP}. Similarly, tools developed with Anthropic’s SDK follow structures such as \textit{@Tool({name: "...", description: "...", ...})}. In such cases, a regular expression parser was used to extract the description string from the relevant function.

Combining the scraped dataset of benign tools with the malicious dataset from the article~\cite{MCPTox} yielded 1,440 unique tool descriptions, each labeled to support both binary and multi-class classification.

\subsubsection{Data Preprocessing}
\label{sec:cleaning_data}
To support the use of scraped data as benign, the tool descriptions were sourced from popular MCP servers managed by trusted developers such as Amazon, Microsoft, and Atlassian, as well as from high-profile repositories. Additionally, the scraped benign data underwent manual review to ensure that all data points were written in English and to verify the absence of any attacks.

To enhance the classification performance of the developed models in real-world scenarios, it was decided to keep the training data as similar as possible to the data the models will retrieve from the live MCP server. Consequently, all stop words were retained, and minimal preprocessing was applied to preserve the original context and meaning of the tool descriptions. Furthermore, during analysis of the malicious dataset, a class labeled “Other,” which contained only two instances, was removed due to its limited contribution to the model’s learning and its tendency to introduce unwanted noise.

\begin{figure*}[b]
    \centering
    \includegraphics[width=1\linewidth]{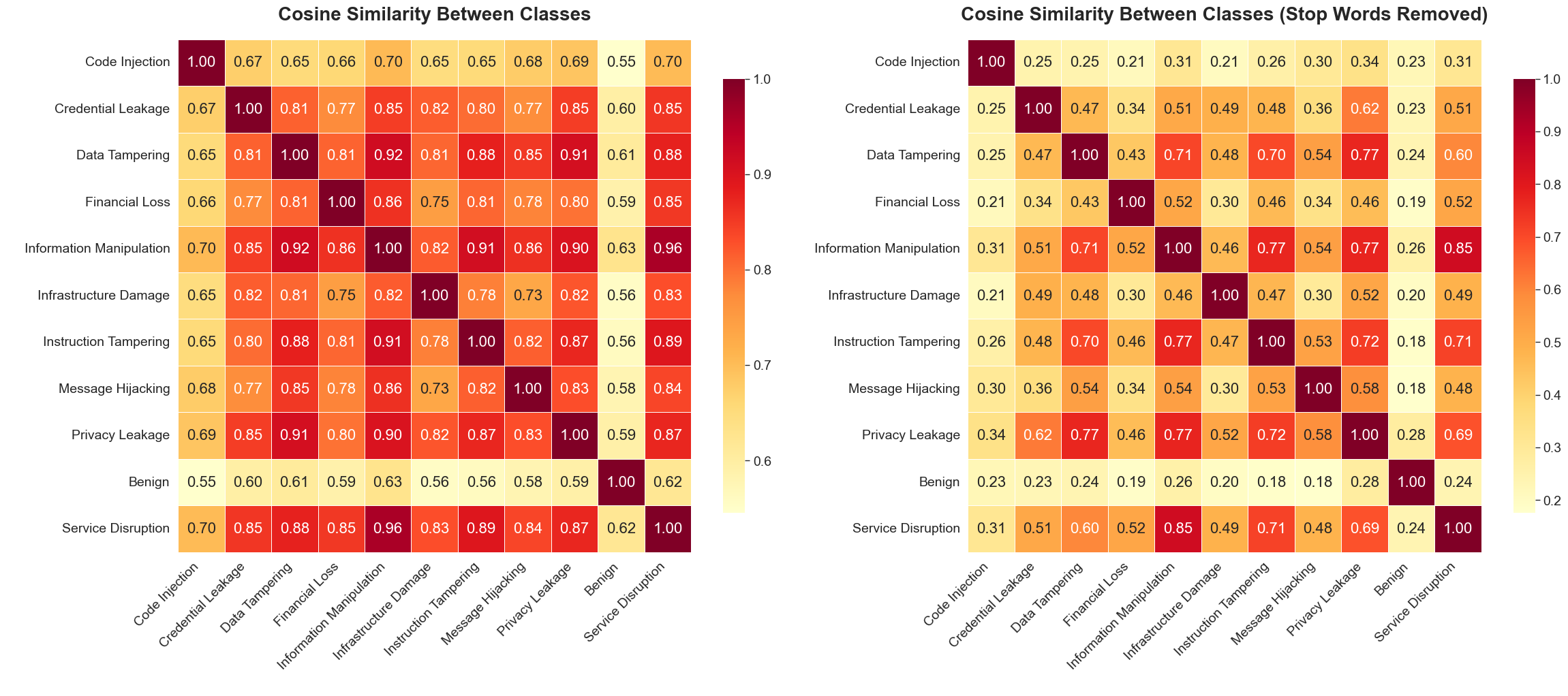}
    \caption{Cosine similarity distance between classes}
    \label{fig:cosinesim}
\end{figure*}

Furthermore, for the multiclass data, class differences were assessed using cosine similarity \cite{SemanticsAngleCosineSimilarity}. As shown in the left panel of Figure \ref{fig:cosinesim}, cosine similarities were computed on the complete, unprocessed raw string tool descriptions. To obtain a more accurate comparison of inter-class similarity, stopwords were removed for this analysis because they disproportionately inflate similarity measures while contributing little to the class-distinguishing content. The right panel of Figure \ref{fig:cosinesim} shows the cosine similarities after stopword removal.

Results from this analysis revealed five classes with low significance differences. The classes — \textit{Service Disruption}, \textit{Privacy Leakage}, \textit{Data Tampering}, \textit{Instruction Tampering} — were merged into a single class labeled \textit{Information manipulation}. This resulted in a new dataset containing seven classes, on which models will be trained, tested, and evaluated. The corresponding cosine similarity matrix showed significantly lower similarity between classes.

\begin{figure}[H]
    \centering
    \includegraphics[width=1\linewidth]{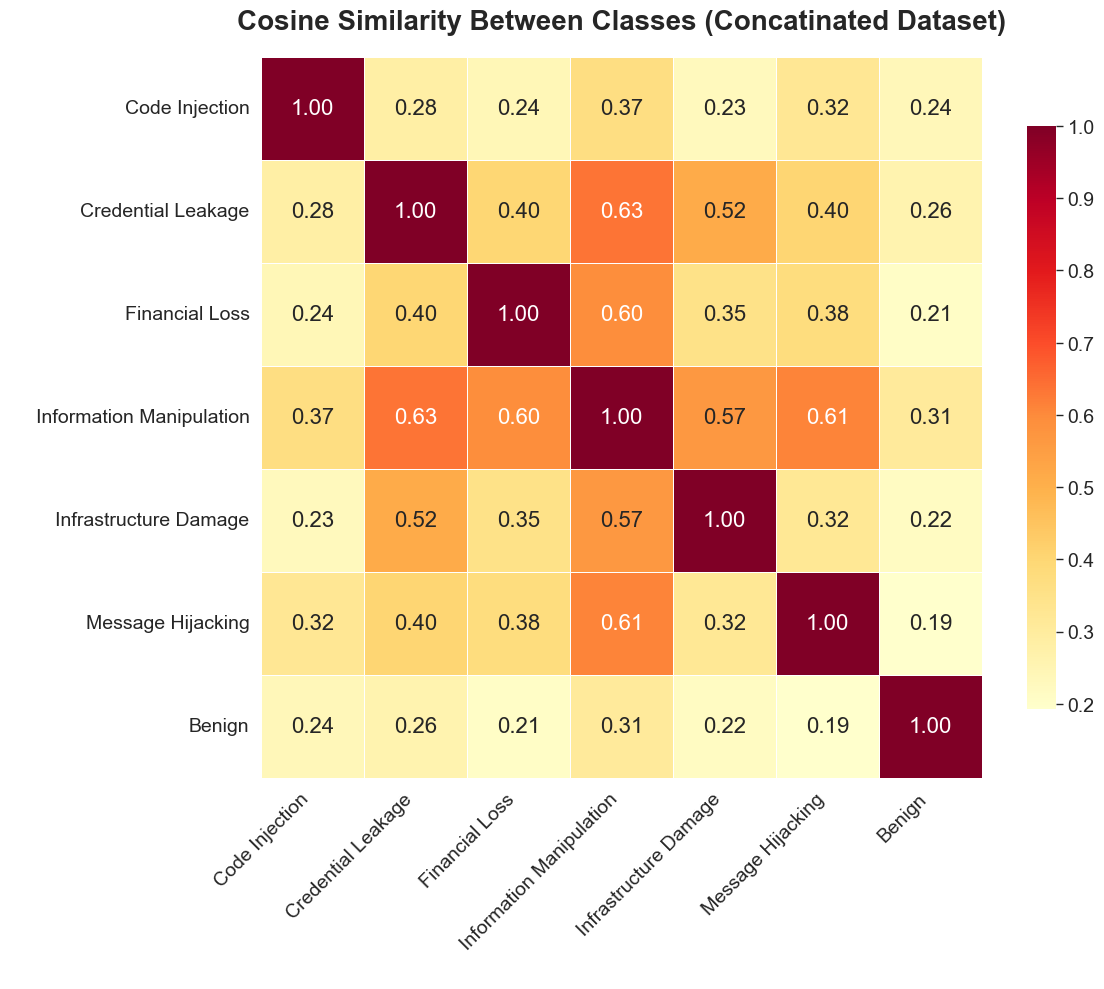}
    \caption{Cosine Similarity matrix after class concatenation}
    \label{fig:cosinesim_red}
\end{figure}

\subsubsection{Expanding dataset}
The initial dataset was insufficient for training. To address this, back-translation was used to augment the dataset~\cite{backtranslation}. Because the project scope was limited to English, languages linguistically distant from English were chosen \cite{LinguisticDistance}: Arabic, Chinese, Japanese, and Korean. For each language, every datapoint was translated into the target language and then back into English, generating paraphrases that preserved the original meaning. To prevent near-duplicates in the final dataset, a similarity filter was applied, and retained only back-translations with a similarity score below 0.8. This process expanded the training set from roughly 1000 data points to nearly 4000.

\subsubsection{Rule based Detection}
Rule-based detection is an essential cornerstone in malware detection and has also been integrated into MCP attack detection systems. Cisco, a large tech company that has integrated YARA rules into its MCP Scanner tool \cite{CiscoMCPScanner}. The YARA rules from this project were used as a benchmark representing the current state of the art. Cisco, however, did not report any performance metrics, so the method’s effectiveness remains unclear. To evaluate the YARA rules, they were imported and applied to our dataset in a controlled test scenario. To illustrate the importance of maintenance, the rules were left unmodified during analysis. The manual nature of this approach is a major drawback: maintainers must regularly update signatures and keywords to keep the system effective, as evidenced by the results. Since YARA rules operate on binary classification, distinguishing only between benign and malicious tool descriptions, they will not be applied to or evaluated for the multi-class classification problem.

\subsubsection{Machine Learning detection}
\label{machine_learning_section}
This study aims to compare machine learning detection approaches with state-of-the-art methods, such as rule-based systems. To identify the best-performing machine learning model for this problem, three different model variants were developed, tested, and evaluated against each other. The selected models, BERT, SVC, and BiLSTM, were chosen based on their demonstrated effectiveness in prior NLP classification tasks \cite{emailSpamDetection}.

\subsubsection{BERT}
\label{bert}
BERT (Bidirectional Encoder Representations from Transformers) is a transformer model introduced by Google in 2018 \cite{devlin2019bertpretrainingdeepbidirectional} and has shown strong performance on a multitude of NLP classification tasks. The base model used in this study, bert-base-uncased \cite{devlin2019bertpretrainingdeepbidirectional}, is pretrained on English and is augmented with a custom classification head: a 10\% dropout layer to reduce overfitting followed by a softmax classifier tailored to the task. The same architecture is used for both binary and multiclass classification; they differ only in the number of output classes in the final layer.

This model is fine-tuned on the dataset using transfer learning and few-shot training \cite{bert_transfer_learning, few_shot_training}. Few-shot fine-tuning has proven effective for leveraging the capabilities of transformer models on small datasets by training the model on task-specific examples for only a few epochs.

\subsubsection{SVC}
\label{svc}
SVC (Support Vector Classifier) is a version of support vector machines used for classification tasks \cite{svm_for_classification}. SVC works by finding a hyperplane in high-dimensional data, making it excellent for NLP tasks that involve sparse or unseen data that differ from the training data. To apply and train an SVC for the detection of tool attacks, the data had to be converted and tokenized, which, for this model, is done using TF-IDF and a hyperplane kernel. For this model, different kernels were tested during early development, but the linear kernel proved to be the best-performing choice for this type of high-dimensional data. \cite{svc_kernels}

\subsubsection{BiLSTM}
\label{bilstm}
BiLSTM (Bidirectional Long Short-Term Memory) is a recurrent neural network that can retain information for long periods, making this architecture perform exceptionally well in many NLP tasks. The bidirectional nature implies that the text sequence is analyzed in both directions, giving the model a better understanding of words in their specific context. BiLSTM was chosen as one of the models applied in this project, primarily because of the similar nature to the field of email spam detection, where LSTM models perform extremely well~\cite{emailSpamDetection}.

\subsection{Solution Evaluation}


To evaluate the capabilities of models presented in Section \ref{machine_learning_section}, two experiments are formulated, the first being binary classification, aligning with the primary goal of any malware defense system, to accurately distinguish benign from malicious data. For each model, we report standard performance metrics computed on held‑out (unseen) samples. The data was split 90-10, with the former used as a train/test data set and the latter set aside as a validation data set. The test dataset was set aside before any data synthesis was performed and kept completely isolated throughout the entire training phase to ensure the validity of the results, thereby minimizing any risk of training data leaking into the test data.

For this study, three models were evaluated: BERT (two variants), SVC, and BiLSTM. The BERT variants comprised of an untuned version that uses only pretrained weights and a tuned version utilizing the technique of transfer-learning and few-shot training \cite{devlin2019bertpretrainingdeepbidirectional} fitted to our classification task. All models were trained and tested on the same data splits. For binary classification only, we also evaluated a YARA rule‑based system; since YARA is designed solely for detection, classifying samples as malicious or benign, and is therefore not suited for multiclass classification.

In the second experiment, multiclass classification was conducted across the different models using the same environment and identical data splits. The only differences between the multiclass and binary classification experiments are the model architectures, multiclass models featuring a different classification head, and the fact that the data in the multiclass experiment is labeled across eleven classes rather than two. The descriptions of the MCP tool used for training and testing remained unchanged between the two experiments.

The experiments were conducted with Pytorch version 2.5.1, Python version 3.10.12, and ran on NVIDIA A100 80GB GPU driver version 550.120 and CUDA version 12.4.

\subsubsection{Evaluation Metrics}
The experimental results were measured using the metrics: accuracy, precision, recall, and F1-score. \\
Accuracy measures the overall proportion of correct predictions and is defined as
\begin{equation}
   Accuracy = \frac{TP + TN}{TP + TN + FP + FN},
\end{equation}
where TP, TN, FP, FN denote True Positive, True Negative, False Positive, and False Negative~\cite{machineLearningTerms}. This provides a high-level summary of model performance but can be misleading when datasets are imbalanced due to the accuracy paradox \cite{accuracyParadox}, which states that in an imbalanced dataset a model can achieve high accuracy simply by predicting the majority class for every sample.

Precision measures how many of the model's positive predictions are actually positive. Precision is measured as:
\begin{equation}
\text{Precision} = \frac{TP}{TP + FP}
\end{equation}
\hspace{0.5cm}
When introducing multiple classes, each class can be treated as the positive class, yielding a binary classification pair for each class and a single precision value per class. To accommodate the additional classes, the precision formula needs to be adjusted. Due to the imbalanced nature of the dataset, this study uses weighted averaging\cite{weightedAveraging}, which is similar to macro-average \cite{metricsForMulticlass}. The weighted averaging precision formula is defined as:
\begin{equation}
\text{Precision}_{\text{weighted}} = \sum_{i=1}^{N} \frac{n_i}{n} \cdot \frac{TP_i}{TP_i + FP_i}
\end{equation}


Recall measures how many positive datapoints the model catches, out of all actual positive labels. \\
Recall is measured as:
\begin{equation}
\text{Recall} = \frac{TP}{TP + FN}
\end{equation}
\hspace{0.5cm}
Similarly to the precision metric, the formula for recall needs to be adjusted in the multiclass scenario, according to the following formula:
\begin{equation}
\text{Recall}_{\text{weighted}} = \sum_{i=1}^{N} \frac{n_i}{n} \cdot \frac{TP_i}{TP_i + FN_i}
\end{equation}

F1-score is the final aggregated metric used in this study; it is the harmonic mean of precision and recall, giving a good balance and a useful measurement when classes are imbalanced. The formula is defined as:
\begin{equation}
\text{F1} = 2 \cdot \frac{\text{Precision} \cdot \text{Recall}}{\text{Precision} + \text{Recall}}
\end{equation}
\hspace{0.5cm}
As the F1-score is based on precision and recall, the formula must be adjusted for the multiclass scenario. F1-scores are computed for each class, weighted by the class's proportion of the total samples, and then summed, resulting in the following formula:
\begin{equation}
\text{F1}_{\text{weighted}} = \sum_{i=1}^{N} \frac{n_i}{n} \cdot 2 \cdot \frac{\text{Precision}_i \cdot \text{Recall}_i}{\text{Precision}_i + \text{Recall}_i}
\end{equation}
\vspace{0.5cm}

To make visualization of the results easier, confusion matrices were added for the different models to show the full distribution of predictions across classes, allowing identification of which classes perform well and which do not. Confusion matrices are especially valuable when analyzing predictions on imbalanced datasets, where the overall accuracy can look impressive, while a confusion matrix can reveal systematic misclassification of the majority class by simply inspecting the off-diagonal elements.

\section{Evaluation Results}
We evaluate the models on two different classification tasks, binary classification and multi-classification with 11 classes, comparing the different models against each other and one state-of-the-art detection method.

\subsection{Binary Classification}
\label{binary_classification_sub}
Based on the results, Table \ref{tab:binary_table} summarizes the metrics obtained. The lowest-scoring model across all five metrics was the untuned (baseline) BERT model. Untuned BERT achieved an accuracy of 56.34\%, compared with 69.01\% for the YARA rule-based detection system. Although these models appear similar when considering accuracy alone, precision, recall, and F1-score tell a different story. YARA achieved precision 62.50\%, recall 20.83\%, and F1-score 31.25\%, whereas the untuned BERT obtained precision 11.11\%, recall 4.17\%, and F1-score 6.06\%.

\begin{figure*}[t]
\centering

\begin{subfigure}{0.3\linewidth}
    \centering
    \includegraphics[width=\linewidth]{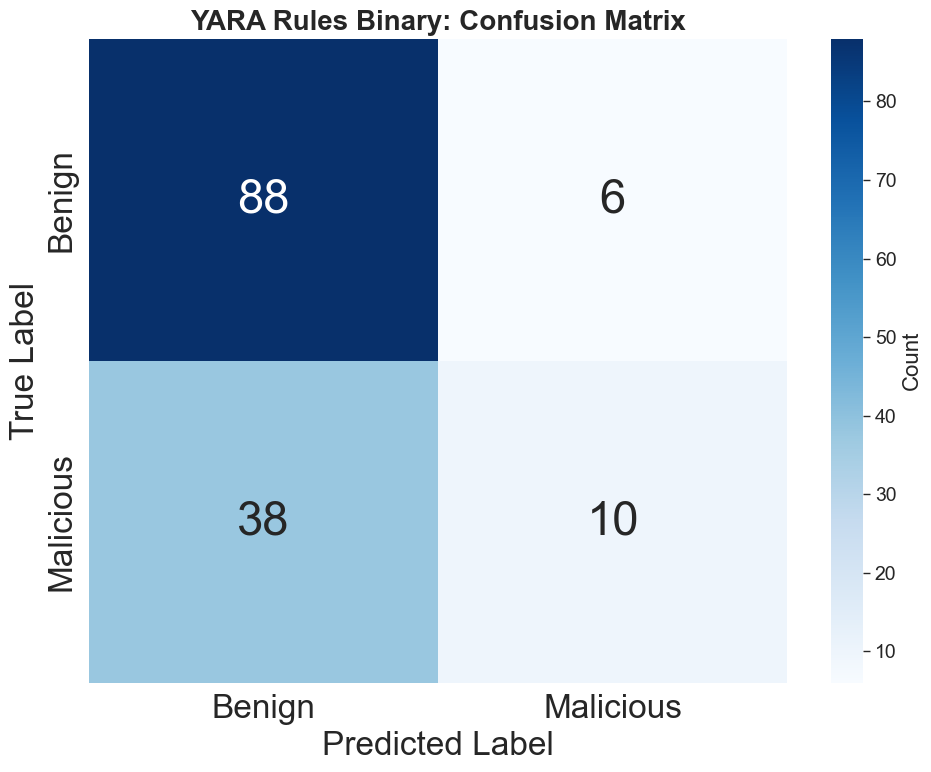}
    \caption{YARA}
    \label{fig:yara_binary_cm}
\end{subfigure}
\hspace{0.5cm}
\begin{subfigure}{0.3\linewidth}
    \centering
    \includegraphics[width=\linewidth]{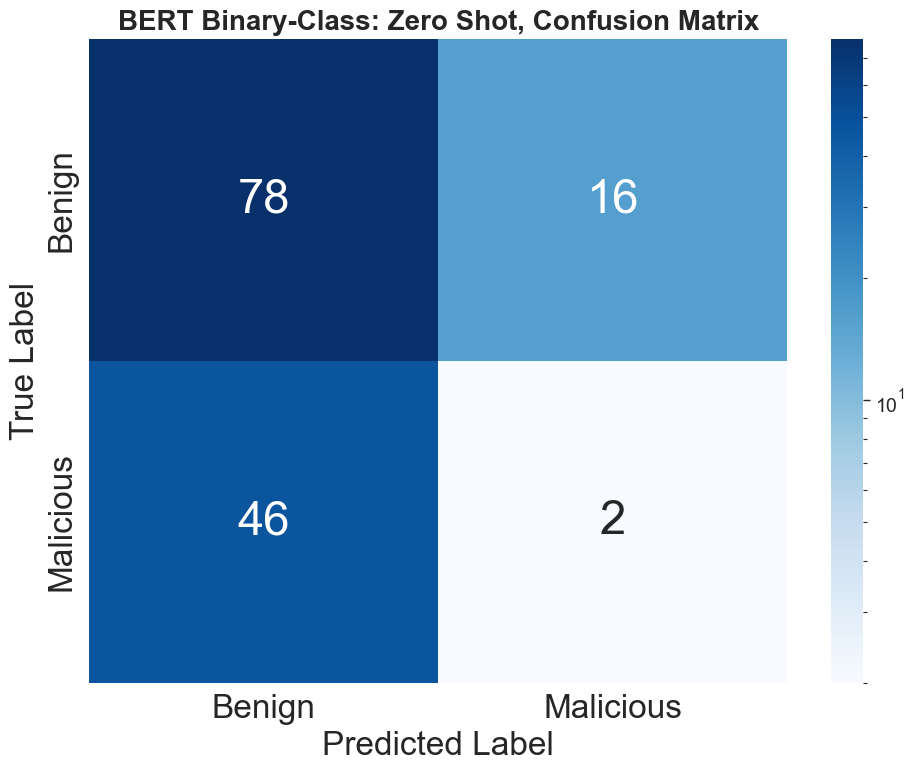}
    \caption{Zero-shot BERT}
    \label{fig:bert_binary_zero_cm}
\end{subfigure}

\vspace{0.5cm}

\begin{subfigure}{0.3\linewidth}
    \centering
    \includegraphics[width=\linewidth]{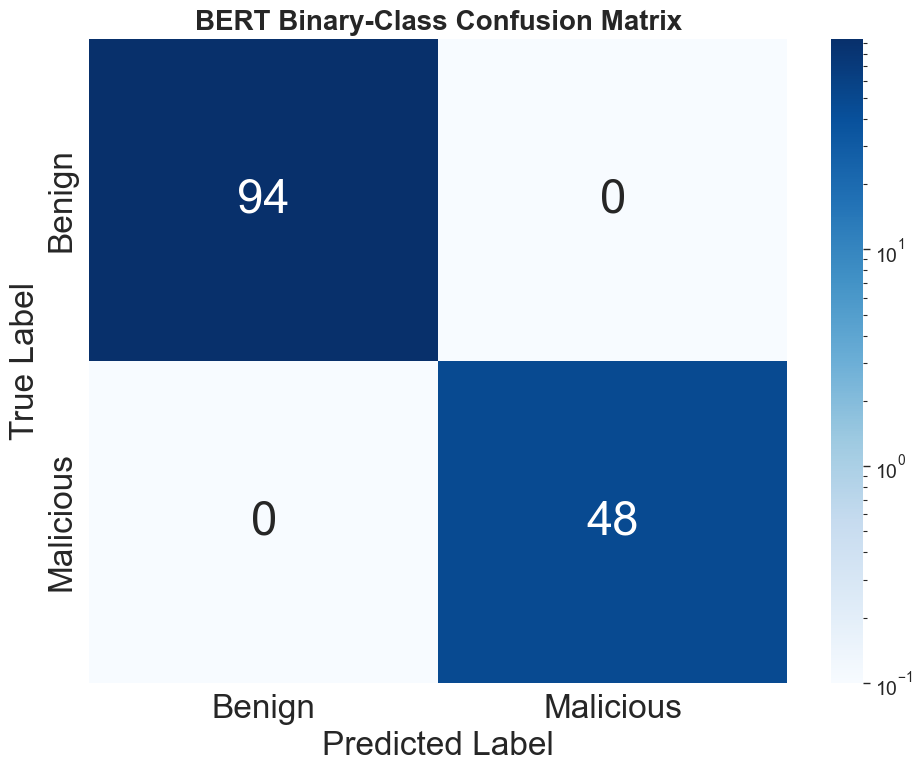}
    \caption{BERT}
    \label{fig:bert_binary_cm}
\end{subfigure}
\hspace{0.5cm}
\begin{subfigure}{0.3\linewidth}
    \centering
    \includegraphics[width=\linewidth]{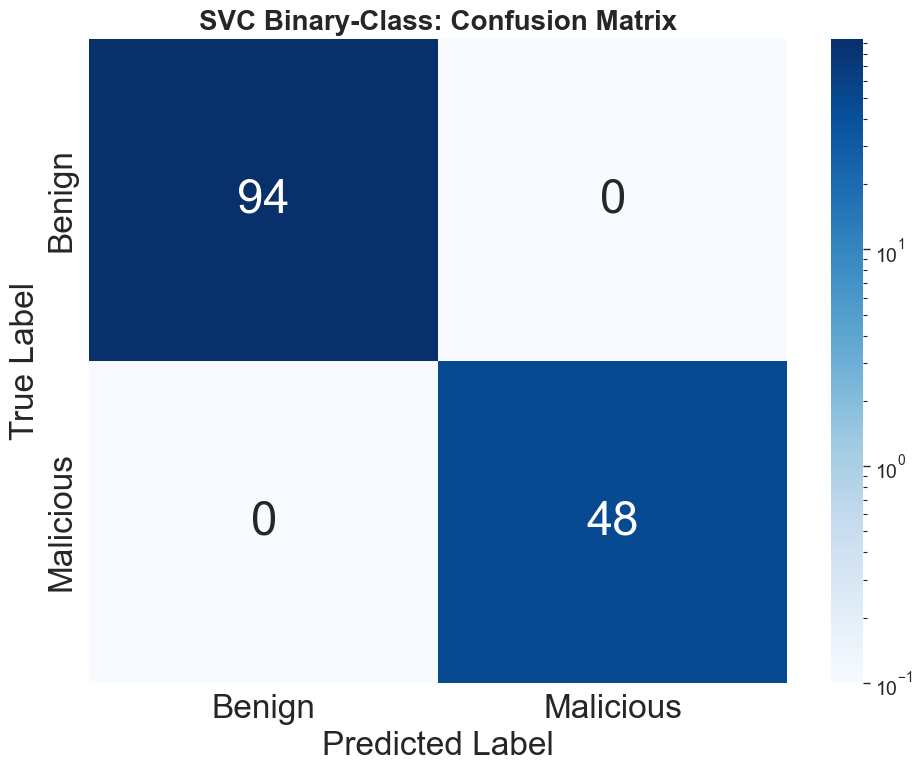}
    \caption{SVC}
    \label{fig:svc_binary_cm}
\end{subfigure}
\hspace{0.5cm}
\begin{subfigure}{0.3\linewidth}
    \centering
    \includegraphics[width=\linewidth]{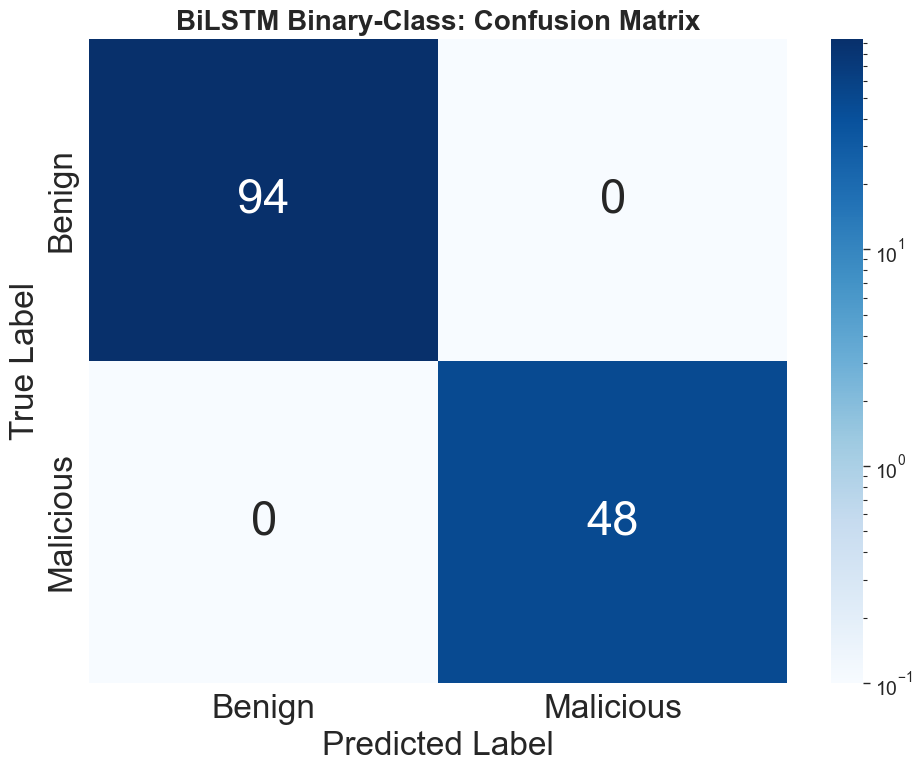}
    \caption{BiLSTM}
    \label{fig:bilstm_binary_cm}
\end{subfigure}

\caption{Confusion matrices for binary classification.}
\label{fig:cm_binary_models}
\end{figure*}

The confusion matrices (Figures \ref{fig:bert_binary_zero_cm} and \ref{fig:yara_binary_cm}) further illustrate this contrast. For the untuned BERT, true positives were 78 out of 94, which appears promising, but true negatives were only 2 out of 48, indicating it rarely correctly classifies malicious tool-description instances, predicting almost all instances as being malicious. YARA performed better in this respect, with 88 true positives out of 94 and 10 true negatives out of 48, though it remained heavily biased towards the benign class.

By contrast, the fine-tuned models (BERT, SVC, and BiLSTM) all produced identical, perfect classification results: accuracy, precision, recall, and F1-score each reached 100\%. Their confusion matrices (Figure \ref{fig:cm_binary_models}) show 94 true negatives and 48 true positives, with no false positives or false negatives.

\begin{table}[H]
    \centering
    \caption{Binary models}
    \label{tab:binary_table}
    \begin{tabular}{lcccc}
        \hline
        Model & Accuracy & Precision & Recall & F1 Score \\
        \hline
        YARA & 69.01\% & 62.50\% & 20.83\% & 31.25\% \\
        BERT * & 56.34\% & 11.11\% & 4.17\% & 6.06\% \\
        BERT & 100.00\% & 100.00\% & 100.00\% & 100.00\% \\
        SVC & 100.00\% & 100.00\% & 100.00\% & 100.00\% \\
        BiLSTM & 100.00\% & 100.00\% & 100.00\% & 100.00\% \\
        \hline
    \end{tabular}
\end{table}

\begin{center}
\textit{* model is untuned only using pre-weights bert-base-uncased}
\end{center}

\subsection{Multiclass Classification}
\label{multiclass}
\begin{figure*}[t]
\centering

\begin{subfigure}{0.4\linewidth}
    \centering
    \includegraphics[width=\linewidth]{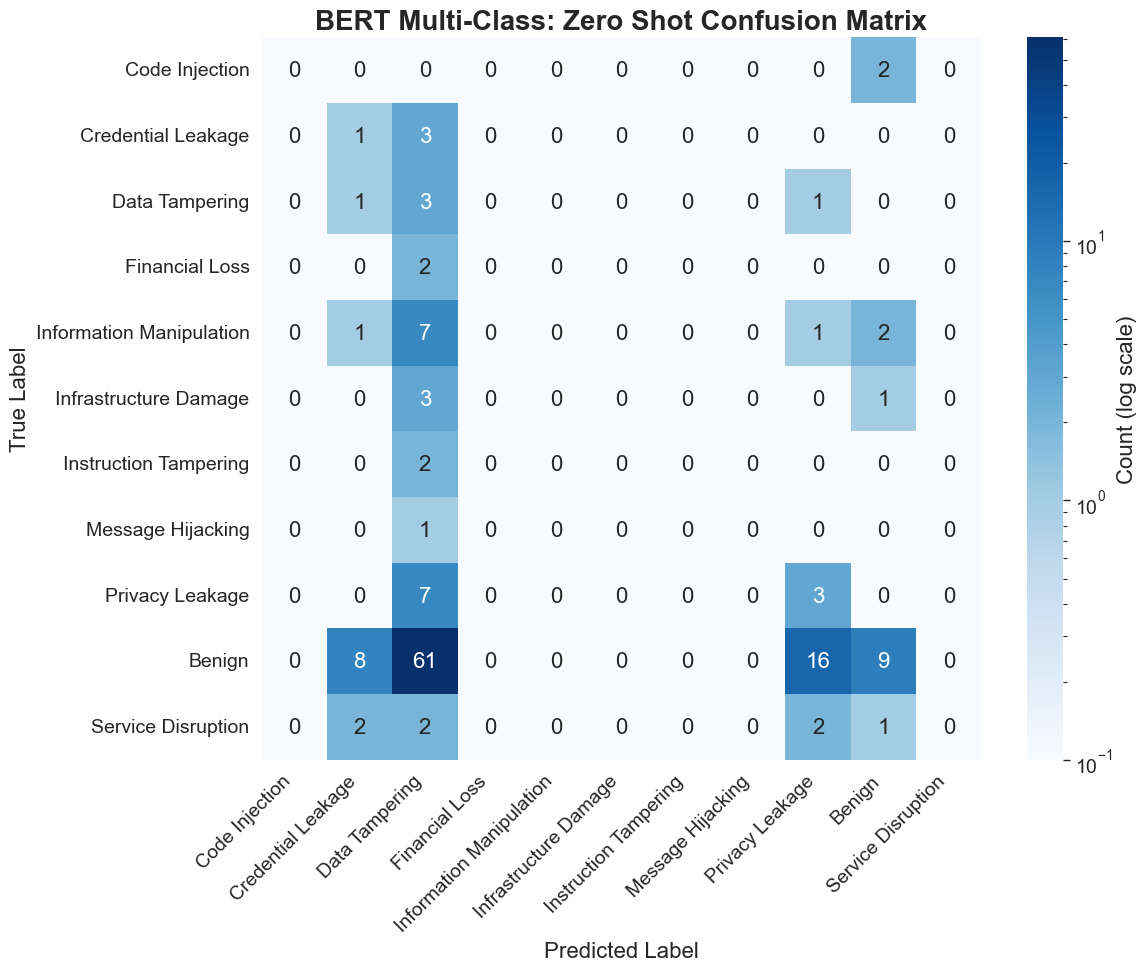}
    \caption{Zero-shot BERT}
    \label{fig:bert_multi_zero_cm_11}
\end{subfigure}
\hspace{0.5cm}
\begin{subfigure}{0.4\linewidth}
    \centering
    \includegraphics[width=\linewidth]{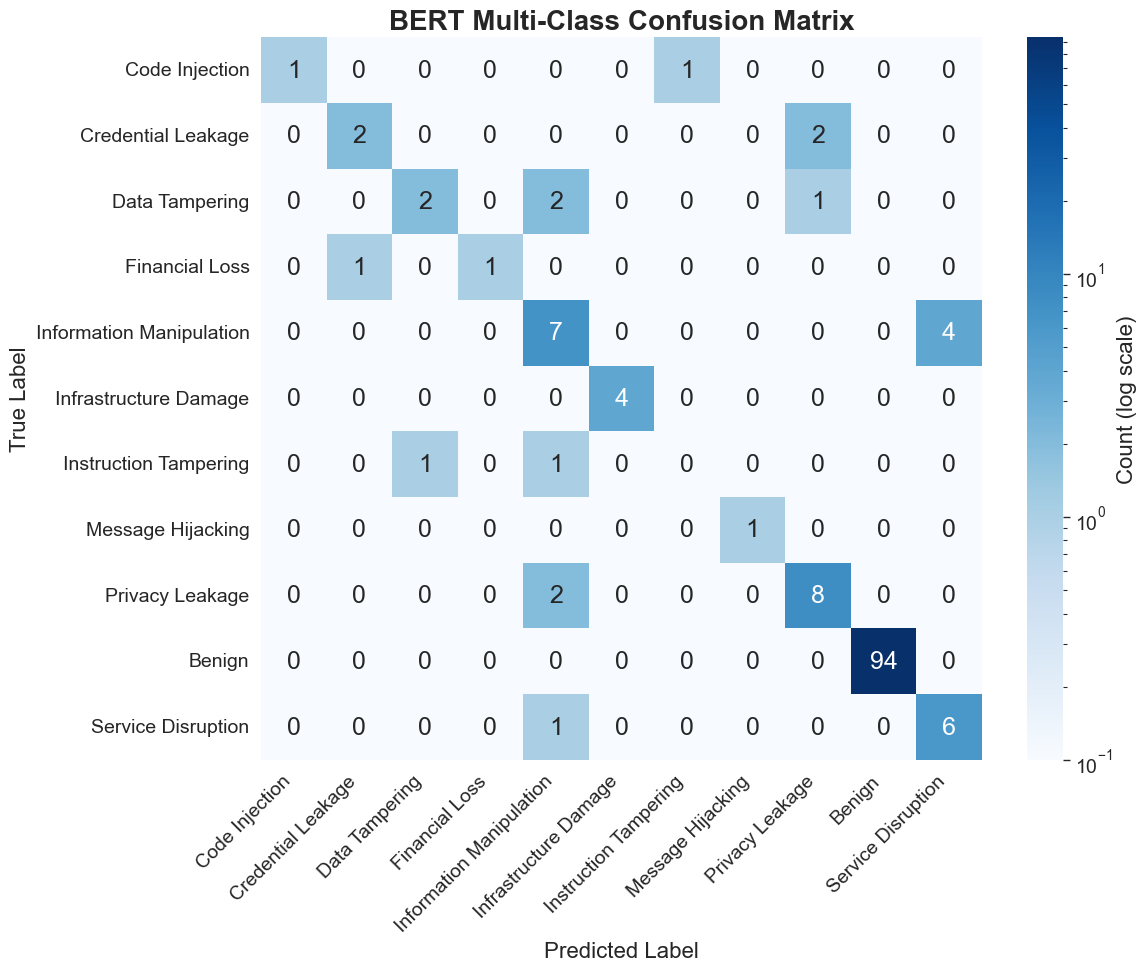}
    \caption{BERT}
    \label{fig:bert_multi_cm_11}
\end{subfigure}

\vspace{0.5cm}

\begin{subfigure}{0.4\linewidth}
    \centering
    \includegraphics[width=\linewidth]{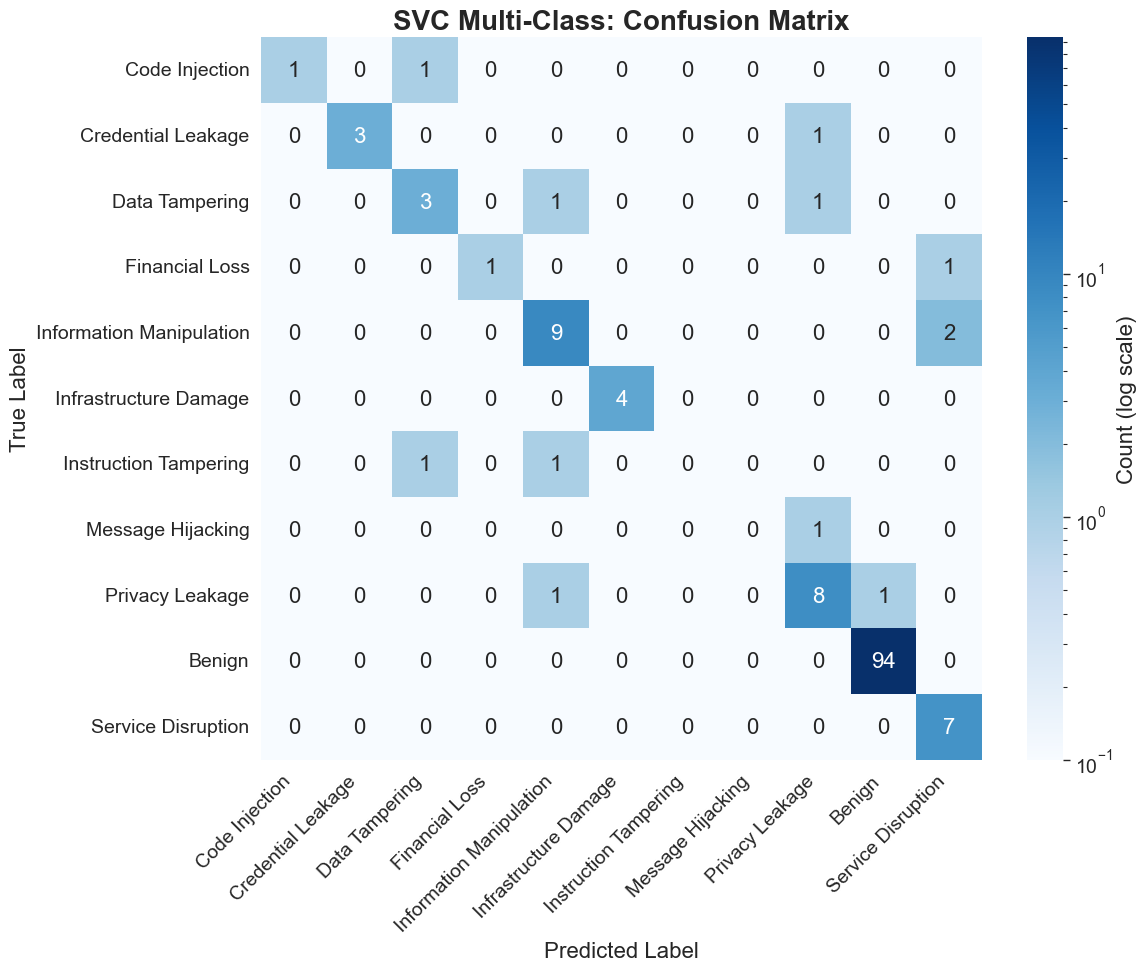}
    \caption{SVC}
    \label{fig:svc_mulit_cm_11}
\end{subfigure}
\hspace{0.5cm}
\begin{subfigure}{0.4\linewidth}
    \centering
    \includegraphics[width=\linewidth]{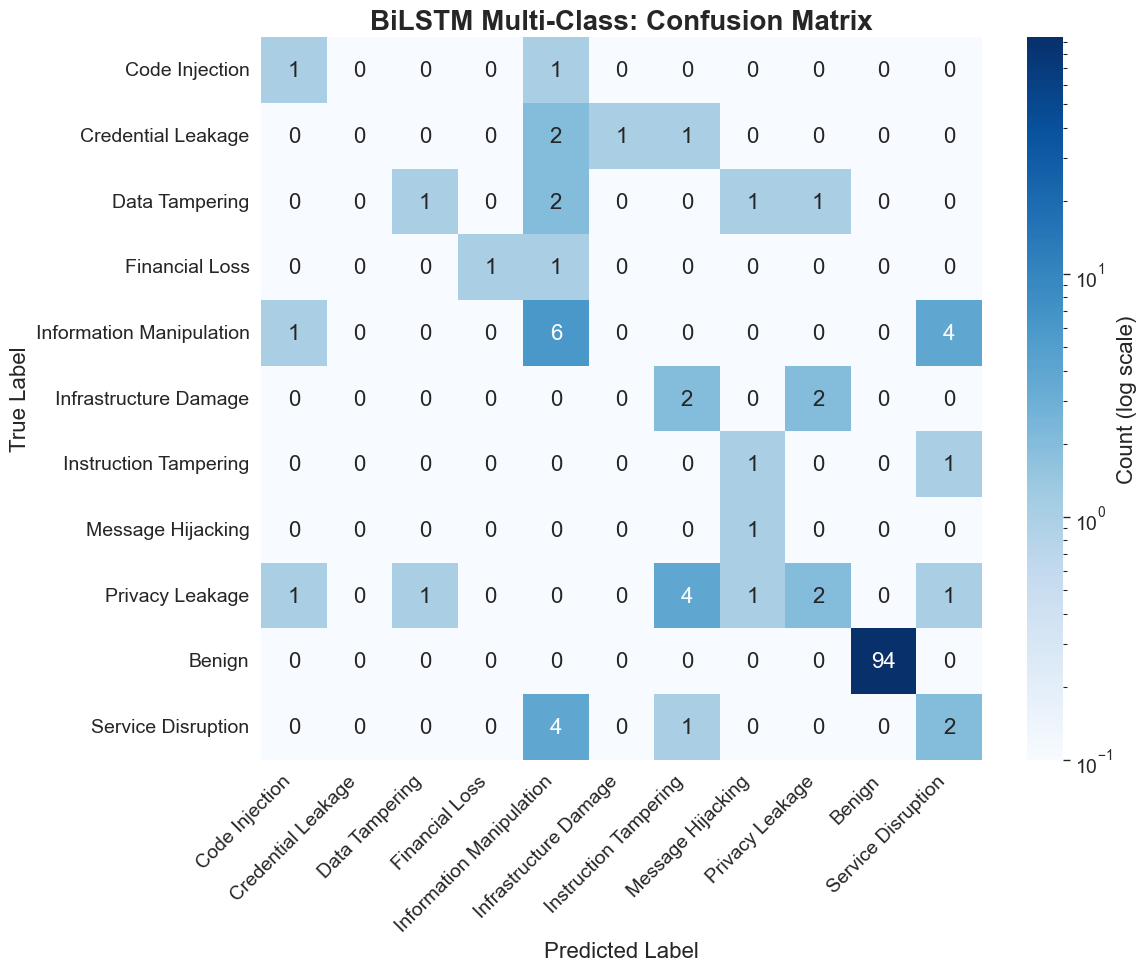}
    \caption{BiLSTM}
    \label{fig:bilstm_multi_cm_11}
\end{subfigure}

\caption{Confusion matrices for multiclass classification.}
\label{fig:cm_multi_models}
\end{figure*}

Multiclass classification, a more complex problem than binary classification, was tested on four models: untuned BERT, fine-tuned BERT, SVC, and BiLSTM. Results presented in Table \ref{tab:multiclass_tab}. Beginning with an untuned model that performs significantly worse than fine-tuned models. Scoring only an accuracy of 11.27\%, Recall 14.79\%, and F1-Score of 12.24\%. Precision appears higher at 40.97\%, but the corresponding confusion matrix (Figure \ref{fig:bert_multi_zero_cm_11}) shows that predictions were concentrated in only four classes. Instead of the ideal diagonal pattern, the confusion matrix exhibits vertical “pillars,” indicating that the model collapsed onto a few classes.

Moving on to the fine-tuned models, achieved substantially higher scores than untuned BERT. SVC performed best overall, with 91.55\% accuracy, 90.45\% precision, 91.55\% recall, and a 90.56\% F1-score. BERT followed closely with 88.73\% accuracy, 89.01\% precision, 88.73\% recall, and an 88.33\% F1-score. BiLSTM achieved the lowest performance among the fine-tuned models, with 76.06\% accuracy, 76.97\% precision, 76.06\% recall, and a 75.62\% F1-score.

The confusion matrices for the fine-tuned models are shown in Figure \ref{fig:cm_multi_models} (SVC: \ref{fig:svc_mulit_cm_11}; BiLSTM: \ref{fig:bilstm_multi_cm_11}; BERT: \ref{fig:bert_multi_cm_11}). BiLSTM shows the most confusion, with a less well-defined diagonal (Figure \ref{fig:bilstm_multi_cm_11}). Despite the greater noise in its predictions, BiLSTM reliably distinguishes benign tool descriptions from malicious ones: it correctly classifies all 94 benign instances as true negatives, yielding no false positives or false negatives for that class. BERT displays a clearer diagonal and likewise correctly classifies all benign instances (Figure \ref{fig:bert_multi_cm_11}). The SVC confusion matrix (Figure \ref{fig:svc_mulit_cm_11}) is similar to BERT but shows an even clearer diagonal; however, one malicious instance with the true label “Privacy leakage” was misclassified as benign.

\begin{table}[H]
    \centering
    \caption{Multi-class models with 11 classes}
    \label{tab:multiclass_tab}
    \begin{tabular}{lcccc}
        \hline
        Model & Accuracy & Precision & Recall & F1 Score \\
        \hline
        BERT * & 11.27\% & 40.97\% & 11.27\% & 12.76\% \\
        BERT & 88.73\%  & 89.01\%  & 88.73\%  & 88.33\% \\
        SVC & 91.55\% & 90.45\% & 91.55\% & 90.56\% \\
        BiLSTM & 76.06\% & 76.97\% & 76.06\% & 75.62\% \\
        \hline
    \end{tabular}
\end{table}

\begin{center}
\textit{* model is untuned only using pre-weights bert-base-uncased}
\end{center}

\subsection{Multiclass Classification with concatenated classes}
During the data-preprocessing stage (see Section \ref{sec:cleaning_data}), it was noted that there was high cosine similarity between classes (Figure \ref{fig:cosinesim}). A new dataset was created by concatenating classes with high similarity, see Figure \ref{fig:cosinesim_red}. Running the multiclass model setup described in Section \ref{multiclass} on this reduced dataset produced the results shown in Table \ref{tab:multi_reduced_table}.

The results are strong, with BERT achieving the best overall performance: 97.89\% accuracy, 98.10\% precision, 97.89\% recall, and a 97.66\% F1-score. SVC achieves 95.77\% accuracy, 95.43\% precision, 95.77\% recall, and 95.25\% F1-score. BiLSTM, similar to the evaluation presented in Table \ref{tab:multiclass_tab}, shows lower performance compared with the two fine-tuned models, with 74.65\% accuracy, 92.77\% precision, 74.65\% recall, and a 74.67\% F1-score. The lowest-performing model was an untuned BERT, which achieved substantially lower scores: 14.79\% accuracy, 45.93\% precision, 14.79\% recall, and 12.24\% F1-score.

\begin{table}[H]
    \centering
    \caption{Multi-class models with concatenated classes}
    \label{tab:multi_reduced_table}
    \begin{tabular}{lcccc}
        \hline
        Model & Accuracy & Precision & Recall & F1 Score \\
        \hline
        BERT * & 14.79\%  & 45.93\%  & 14.79\%  & 12.24\% \\
        BERT  & 97.89\%  & 98.10\%  & 97.89\%  & 97.66\% \\
        SVC & 95.77\%  & 95.43\%  & 95.77\%  & 95.25\% \\
        BiLSTM  & 74.65\%  & 92.77\%  & 74.65\%  & 74.67\% \\
        \hline
    \end{tabular}
\end{table}

\begin{center}
\textit{* model is untuned only using pre-weights bert-base-uncased}
\end{center}

\subsection{Token Importance Analysis}
\label{token_importance}

Figure \ref{fig:bert_token_bias} consists of two panels showing the top 40 most influential tokens for each class in the binary classification setting. Tokens with negative values indicate which tokens the model associates with benign behavior, and are displayed in the left panel. Similarly, the right panel displays the tokens with the highest positive values, which the model associates with malicious behavior. The token bias calculation uses gradient-based attribution\cite{integratedGradients} to measure each token's contribution to the classification decision. For each token, it gets multiplied by its embedding value, and the value is then summed over all dimensions. The result is a single influence score displayed in the bar plot panels.

\section{Discussion}
The aim of this study was to develop a machine learning model that could efficiently distinguish between benign and malicious MCP tool descriptions. The findings reveal an overall promising performance in both classification tasks. Beginning with binary classification — distinguishing between benign and malicious tools, the core task of any malware-detection system. The results, show in Figure \ref{tab:binary_table} indicate promising overall performance. 
However, because the dataset is imbalanced to reflect real-world conditions (there are more benign than malicious samples), some models and systems—specifically the rule-based YARA system and the untuned BERT*—obtain inflated scores by predominantly predicting samples as benign. When considering only accuracy, these models still appear effective, but metrics such as precision, recall, and F1-score indicate limited performance. The poor ability of these approaches to detect malicious samples is further evident in the confusion matrices (\ref{fig:bert_binary_zero_cm} and \ref{fig:yara_binary_cm}), which clearly illustrate this behavior.

On paper, YARA rules can effectively detect certain malicious instances and perform well on other tasks, including MCP descriptions (see \cite{automatedYaraRules}). However, in settings with unseen MCP tool descriptions, where the rule-based system has not been tuned for specific attack terms and where context matters more than individual words, classification becomes too complex for such a system to handle accurately. In this instance, YARA is unable to classify with any practical performance. The same can be seen in the untuned version of BERT, where the models allocate most of their predictions to benign tool descriptions.

After analyzing the two poorly performing systems, the trained models BERT, SVC, and BiLSTM demonstrate much stronger performance in the binary classification task, with all three achieving perfect accuracy and correctly classifying every example. This result likely indicates that these trained models have learned contextual cues that distinguish malicious from benign descriptions—something the untuned BERT model appears unable to do. A perfect score may also suggest that the binary classification problem is relatively trivial for advanced models. Referring back to the class similarity plot in Figure \ref{fig:cosinesim}, a clear separation can be observed between the benign class and all malicious classes, a distinction that BERT, SVC, and BiLSTM seem able to capture effectively.

Moving on to the more complex multiclass classification task for MCP-tool description attacks, which was added as an extension after we observed promising initial results in the binary classification task (Section \ref{binary_classification_sub}). This task uses the full labeling from the dataset developed in \cite{MCPTox}, consisting of 11 classes: 10 malicious and 1 benign. To keep the experiments comparable to the binary classification setup, we modified only the models’ classification heads to output eleven classes instead of two; other parameters (optimizers, loss functions, etc.) remained unchanged.

Results from the multiclass experiments are generally strong, except for the untrained version of BERT. This outcome was expected based on the binary classification results (Section \ref{binary_classification_sub}): Increasing the number of classes and the problem’s dimensionality further reduced performance, likely because the model struggled to capture contextual nuances and determine how to utilize the given data. As a result, patterns similar to those shown in Figure \ref{fig:bert_multi_zero_cm_11} emerge, where the classes collapse into each other, ineffectively stacking everything into four classes.

Among the trained and hyperparameter-tuned models, the BiLSTM shows promise in distinguishing malicious classes. Achieving an accuracy of 74.65\% and a precision of 92.77\%. The confusion matrix (Figure \ref{fig:bilstm_multi_cm_11}) indicates that although the model struggles with noisy data, probably because many classes have high contextual similarity, it still perfectly separates benign and malicious instances (100\% binary accuracy). This demonstrates that, despite difficulties in fine-grained multiclass labeling, the BiLSTM preserves the core functionality of malware detection: distinguishing attacks from non-attacks.

Continuing, BERT and SVC achieve impressive performance in multiclass classification of MCP-tool descriptions. Both models can, to a large extent, distinguish among different malicious classes while maintaining nearly perfect binary separation between benign and malicious instances. The only benign/malicious misclassifications occurred with SVC, which labeled a privacy-leakage instance as benign, whereas BERT retained 100\% binary accuracy even on multi-labeled data. BERT and SVC exhibit similar metric values in Table \ref{tab:multiclass_tab}. Examination of their confusion matrices (Figures \ref{fig:bert_multi_cm_11} and \ref{fig:svc_mulit_cm_11}) reveals a strong diagonal structure, indicating excellent multiclass performance.

Related to the BERT model, the token importance analysis (see Figure~\ref{fig:bert_token_bias}) shows which tokens influence the model’s predictions. The benign-biased tokens are mostly related to data retrieval keywords such as \textit{about}, \textit{get}, and \textit{details}. It also indicates some tendencies toward structuring operations, like \textit{list}, and \textit{group}.

For the malicious-biased tokens, on the other hand, the analysis reveals a lot of instructional language such as \textit{must}, \textit{should}, \textit{will}, suggesting that the model has learned that malicious tool descriptions frequently contain prescriptive or step-by-step instructions. Interestingly, tokens like \textit{delete} or \textit{execute} were not among the most influential malicious-biased tokens. This could also explain why rule-based solutions like Yara do not perform as well as models that actually learn the context of the tool descriptions.

Further, after seeing the cosine similarity analysis of the multi-labeled dataset described in Section \ref{sec:cleaning_data}, we observed that five classes remained highly similar even after stopwords were removed. These classes were concatenated into a single class, yielding a dataset with seven classes, where they now have reduced cosine similarity (see Figure \ref{fig:cosinesim_red}). Training and evaluating the models on this reduced dataset revealed performance improvements (see Table \ref{tab:multi_reduced_table}). BERT achieved an impressive accuracy of 97.89\% and a precision of 98.10\% compared to previous accuracy of 88.73\% and precision of 89.01\% (see Table \ref{tab:multiclass_tab}). An improvement with around 10\%. Although these results are substantially better than those obtained on the original eleven-class problem discussed in Section \ref{multiclass}, the classification space was reduced by four classes, effectively removing noise by grouping similar classes together and simplifying the task. There is, however, a trade-off between reducing dimensionality and preserving a genuine multiclass problem: merging too many classes risks reverting the problem closer to a binary classification. Therefore, for this study, we retained the original eleven classes.

While this study shows great potential for detecting malicious MCP tools, with impressive results in the binary classification task, it also has certain limitations, such as analyzing only MCP tools with English-language descriptions. One reason is that most MCP tools are developed in English. Furthermore, the pretrained BERT model and the embeddings model used for data processing in the BiLSTM are exclusively trained on English. Further, the research team for this study does not understand other languages that could be relevant, such as mandarin, a language where there exist a large quantity of MCP tools descriptions. But we would not be able to manually control the validity of the content descriptions due to the language barrier.

Additionally, since this study focuses on analyzing natural language data, the project is also limited to the tool docstrings descriptions of MCP tools. This means that if the code associated with a tool contains harmful syntax, this detection tool will not detect it. Another reason for this limitation is that, at the time of the study, most MCP injection attacks target tool descriptions due to the universal tool support across MCP clients\cite{MCPClientSupport}.

Furthermore, limitations to the validity of this research include limitations in the data, since the training data were collected in two different ways, as mentioned in Section \ref{sec:aquiring_data}. The malicious data points were collected solely from the MCPTox Library \cite{MCPTox}. This limits the trained model's perception exclusively to the data included in MCPTox. Besides, the classes used in the multiclass classification are based on the threat classes presented in the MCPTox dataset. Consequently, if a malicious MCP tool has a description that differs substantially from the training data, the detection model—especially in the multiclass setting—will struggle to detect it. The malicious data for this study was built on the MCPTox library, and benign data points were collected from publicly available MCP servers. To verify that the tool descriptions from the public MCP servers were actually benign, we manually reviewed each one. Because human error is a non-negligible risk, some mislabeled data may be present.

\section{Conclusions and Future Work}
This section presents the key findings obtained in this study on the detection of malicious MCP tools using deep learning methods. The main results are discussed in relation to the initial research goals, demonstrating the effectiveness of the proposed models. In addition, the limitations encountered during the research are considered, and recommendations are provided for extending this work in future studies.

To address the lack of security measures against MCP tools, this study presents a novel method that uses multiple machine learning algorithms to assess the safety of these tools. By connecting the tool to active servers hosting MCP tools, malicious instances can be detected and avoided before any harm occurs.

In addition to the final model integrated into the tool, three supplementary models were developed and compared to ensure that the most effective approach was applied to this problem. The findings demonstrate that advanced machine learning models are highly capable of detecting malicious MCP tool descriptions—particularly BERT and SVC, which performed well in both binary and multiclass classification tasks by interpreting and classifying text based on context.

Furthermore, the models developed in this research outperformed existing rule-based systems, such as YARA rules, which primarily rely on keyword matching. Building on this observation, the results demonstrate that there are no definitive features in MCP-tool descriptions that directly indicate classification, as shown in Section \ref{token_importance}. Rather, the overall context of the tool description determines whether a tool is harmful or benign. These findings contradicted our initial hypothesis at the outset of this work, which assumed that machine learning would serve as an additional layer on top of existing methods to enhance detection. Instead, the evidence indicates that there is no need to combine rule-based techniques with our machine-learning–based detection, as the trained models have surpassed current approaches by a substantial margin. This superior performance renders existing methods, such as YARA rules, largely obsolete.

During the development phase of this project, we noted that Anthropic has released a new workflow called Claude Skills \cite{agent_skills}, which extends MCP's agentic capabilities in a manner similar to MCP tools. Similar to MCP tools, Skills consist of data built using natural language. At the time of this study, Skills suffered from the same lack of guardrails as MCP tools, and its attack surface appeared even larger. 
The development of a detection tool for Claude Skills is highly relevant, and we have attempted to apply our scan tool to it. Even though it showed some effectiveness, it became clear that skill-specific data is necessary to train a truly effective detection model, which is outside the scope of this project.


\setlength{\bibsep}{4pt}
\bibliography{thesis-refs}
\bibliographystyle{ieeetr}

\clearpage
\onecolumn
\appendix
\section{Appendix}

\begin{figure}[h]
    \centering
    \includegraphics[width=\linewidth]{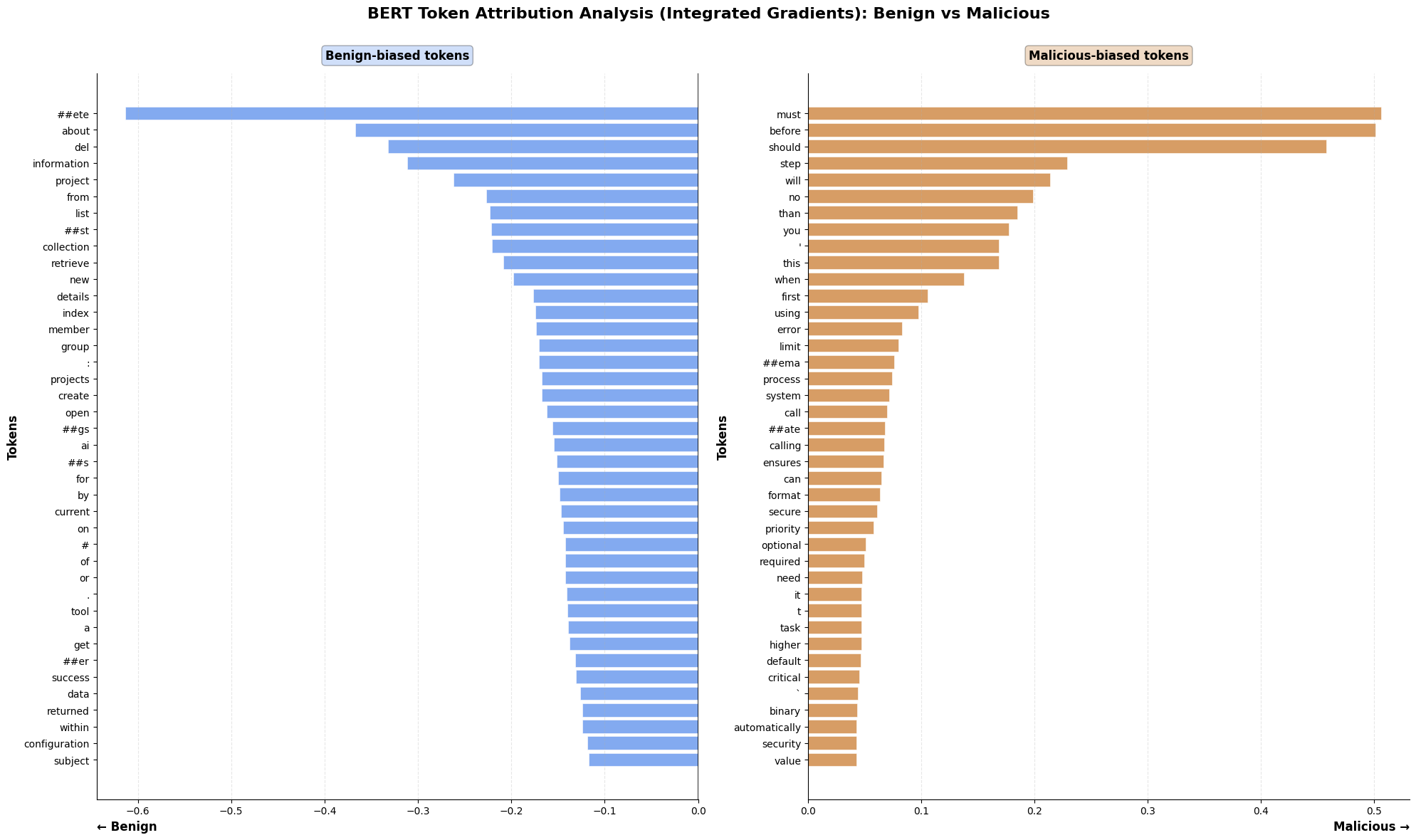}
    \caption{Visualization of the tokens that influence the BERT model's decisions.}
    \label{fig:bert_token_bias}
\end{figure}

\end{document}